# Giant Anisotropic in-Plane Thermal Conduction Induced by Anomalous Phonons in Nearly-Equilaterally Structured PdSe$_2$

Bin Wei[1,2*], Junyan Liu[2*], Qingan Cai[3], Ahmet Alatas[4], Ayman H. said[4], Chen Li[3,5†], Jiawang Hong[2†]

[1]Henan Key Laboratory of Materials on Deep-Earth Engineering, School of Materials Science and Engineering, Henan Polytechnic University, Jiaozuo 454000, China

[2]School of Aerospace Engineering, Beijing Institute of Technology, Beijing 100081, China

[3]Mechanical Engineering, University of California, Riverside, Riverside, CA 92521, USA.

[4]Advanced Photon Source, Argonne National Laboratory, Argonne, IL 60439, USA.

[5]Materials Science and Engineering, University of California, Riverside, Riverside, CA 92521, USA.

*The authors contribute equally

†Corresponding author: chenli@ucr.edu.cn; hongjw@bit.edu.cn

## Abstract

In two-dimensional materials, structure difference induces the difference in phonon dispersions, leading to the anisotropy of in-plane thermal transport. Here, we report an exceptional case in layered PdSe$_2$, where the bonding, force constants, and lattice constants are nearly-equal along the in-plane crystallographic axis directions. The phonon dispersions show significant differences between the Γ–X and Γ–Y directions, leading to the anisotropy of in-plane thermal conductivity with a ratio up to 1.8. Such anisotropy is not only unexpected in equilaterally structured (in-plane) materials but also comparable to the record in the non-equilaterally structured material reported to date. By combining inelastic X-ray scattering and first-principles calculations, we attribute such anisotropy to the low-energy phonons along Γ–X, in particular, their lower group velocities and

"avoided-crossing" behavior. The different bucking structures between *a*- (zigzag-type) and *b*-axis (flat-type) are mainly responsible for the unique phonon dynamics properties of PdSe$_2$. The present results illustrate the unusual thermal conduction mechanism of the equilaterally structured materials and provide valuable insights on thermal management in electronic devices.

## I. INTRODUCTION

Due to their excellent physical properties, low-dimensional materials play an irreplaceable role in meeting the requirements of intelligent and miniaturized devices. Potential applications of these materials have been studied in a variety of fields, such as electronics [1-3], sensors [4, 5], and spintronics [6, 7]. However, controlling heat dissipation remains a challenge, making thermal management an intriguing topic in optimizing low-dimensional materials. In two-dimensional (2D) layered materials, the weak interlayer van der Waals (vdW) force and the strong intralayer covalent bonds lead to highly anisotropic thermal conduction between the out-of-plane and in-plane directions. Not surprisingly, the anisotropy of in-plane thermal conductivity (IPTC) also exists in many 2D materials [8-11]. For example, in a hexagonal structure system (such as graphene [8] and black phosphorus [9]), the thermal conductivities along the crystallographic axes are equal due to the planar honeycomb structure, while the IPTC along the zigzag- and armchair-directions are anisotropic in the basal plane. In the non-equilaterally structured titanium trisulfide, a high anisotropy ratio of 2 has been reported due to the large difference of bonding and lattice constants between along *a*- and *b*-axis [11]. Therefore, it is a meaningful question to study whether the IPTC anisotropy exists in the equilateral in-plane structure, which may open up a new way for optimizing device performance. Recently, pentagonal-structure 2D materials, such as penta-graphene [12], penta-silicene [13], and noble metal dichalcogenides (MX$_2$) (M = Pt, Pd, and X = S, Se, Te) [14], have become the focus of interest due to their unique structures. Among them, palladium diselenide ($PdSe_2$) has attracted extensive attention, thanks to its unique electrical, mechanical, and optical properties [15-18]. Anisotropy of $PdSe_2$ has also been reported, such as electrical [19, 20], optical [15, 19], and thermoelectric properties [16, 21]. Much work has focused on the theoretical lattice dynamics calculations, but less was confirmed by measurements [22-24]. $PdSe_2$ takes a puckered pentagonal structure with an orthorhombic lattice ($a$=5.75 Å, $b$=5.87 Å, $c$=7.69 Å, $a/b$=0.98) (Figure 1a) [25], in which four Se atoms covalently bond the Pd atom with a nearly-equal Pd-Se bond length (2.44 Å / 2.44 Å ≈ 1), bond strength (2.51 eV /2.50 eV ≈ 1), and bond force constant (4.18 eV Å$^{-2}$ / 4.15 eV Å$^{-2}$ ≈ 1) [26], showing a nearly-isotropic structure in the basal plane (Figure 1b). It is significantly different from the equilateral structure pentagonal materials [12, 13, 27]: well-studied graphene structure with the zigzag and armchair directions [28], or the $MX_3$ (M = Zr, Ti, and X = S, Se) with the one-dimensional chain direction [29]. Therefore, it is interesting to explore the thermodynamics in such peculiar structure. Recently, an anisotropy of IPTC with a ratio up to 1.4 has been reported by micro-Raman thermometry (MRT) [25]. However, the microscopic mechanism of this thermal conductivity anisotropy, especially the lattice dynamics, is still unknown experimentally.

Here, we report a giant anisotropy of IPTC with a ratio up to ~ 1.8 in layered $PdSe_2$ with the nearly-equilateral structure over a wide temperature range. By using inelastic X-ray scattering (IXS) and first-principles calculations, we find that such anisotropic thermal conduction is induced by flat low-energy (LE) phonons dispersion and avoided-crossing behavior along Γ–X, both of which strongly suppress the heat transport. This unique transport behavior is mainly induced by the different bucking structures between $a$- (zigzag-type) and $b$-axis (flat-type) in $PdSe_2$ (Figures 1c and 1d).

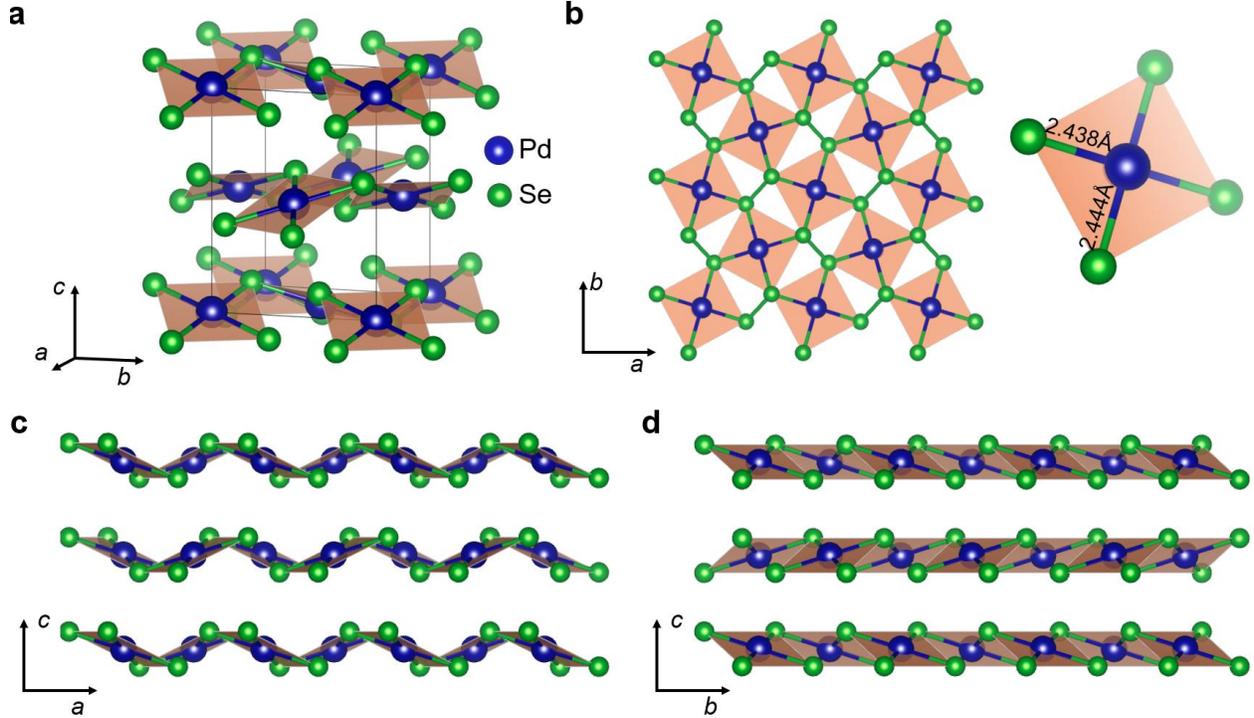

Figure 1. Crystal structure of PdSe$_2$. **a** and **b** are the front and top views of the crystal structure, respectively. The nearly-equal Pd-Se bond lengths showing nearly-regular quadrilaterals. **c** and **d** are the side views, showing a zigzag-type pucker along *a*-axis and a flat-type pucker along *a*-axis, respectively.

## II. EXPERIMENT AND CALCULATION

### Inelastic X-ray scattering

2D-layered PdSe$_2$ crystals in this work were grown by the CVT method. The quality of crystals was checked by X-ray diffraction shown in Figure S1 in the supporting information (SI). The full width at half maximum (FWHM) of the X-ray diffraction peak at (002) plane is 0.24 ± 0.01°, indicating high-quality. The LE phonon dispersions were measured at room temperature by IXS experiment conducted at 30-ID-C (the High-resolution Inelastic X-ray Scattering beamline, HERIX) at the Advanced Photon Source (APS) [30, 31]. A single crystal of PdSe$_2$ with 60 μm thickness was attached to a copper post by varnish (inset in Figure S1). The incident photon energy was 23.7 keV with an energy resolution Δ*E* of 1.2 meV. The IXS measurements were accomplished at constant wavevector mode in transmission geometry. The orientation matrix was defined by using Bragg peaks at (2 0 0), (0 2 0), and (0 0 2).

**Calculations**

First-principles calculations were performed based on the density functional theory (DFT) in the projector-augmented wave (PAW) [32] framework as implemented in the Vienna Ab Initio Simulation Package (VASP) [33]. The generalized gradient approximation (GGA) of Perdew-Burke-Ernzerhof (PBE) to the exchange-correlation functional was used, and the energy cut-off of 600 eV was set for the plane-wave basis cut-off energy of 600 eV. We used optPBE functional [34] to evaluate the vdW force due to layered structure $PdSe_2$. The Brillouin zone of the reciprocal space is sampling by a $\Gamma$-centered grid of 5 ×5 × 4. The force components of each atom are smaller than 0.001 eV/Å, and the difference of total energy was less than $10^{-6}$ eV during relaxing the atomic positions and electrons. The lattice constants used for calculations are from IXS measurements, where a = 5.741 Å, b = 5.868 Å, and c = 7.705 Å.

The Phonopy code [35] was used to calculate the phonon dispersion of layered $PdSe_2$. In the approach, the second-order interatomic force constants (IFCs) were computed by the finite difference method in a 2 × 2 × 2 supercell. The phonon transport properties were obtained by solving the phonon Boltzmann transport equation (BTE) implemented in the ShengBTE package [36]. The four neighbor interactions were considered to capture the basic phonon scattering processes embedded in this software to calculate the anharmonic IFC. 10 × 1 × 1, 4 × 1 × 1 and 10 × 1 × 1 supercell containing 120, 48, and 120 atoms are used to map out the mode potential energy surfaces along $\Gamma$–X and $\Gamma$–Y directions at $q$ = 0.1, 0.25 and 0.3, respectively.

**III. RESULTS AND DISCUSSION**

Figure 2 shows the LE phonon dispersions of PdSe$_2$ measured by IXS at room temperature along Γ–X, Γ–Y, and Γ–Z directions, overlaid with the calculated dispersions (the entire energy dispersions are shown in Figure S2). It can be seen that the calculation is in excellent agreement with IXS measurements, in particular along the interlay direction along vdW forces (Figure 2c). The accurate modeling of the phonon dispersion is essential for thermal conductivity calculations. Interestingly, the LE phonon branches along Γ–X show significant differences from those along Γ–Y.

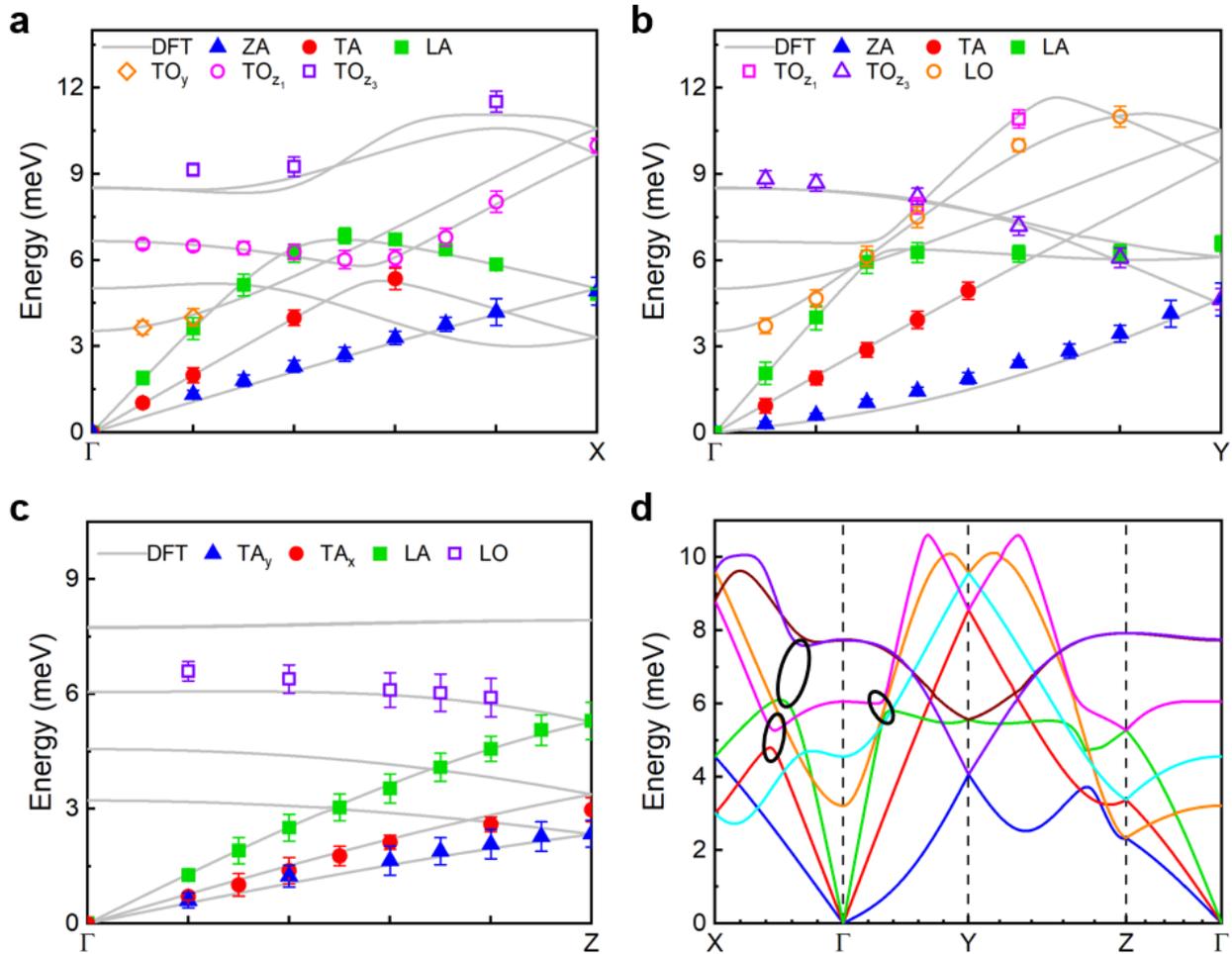

Figure 2. Phonon dispersions of PdSe$_2$ measured at 300 K by IXS, overlaid onto the calculations. Points are the measurements, and solid lines are the calculations. **a**, **b**, and **c** are the dispersions along Γ–X, Γ–Y, and Γ–Z, respectively. (solid lines). Error bars are the fitting uncertainties. **d** Calculated LE phonon dispersions, sorted by the phonon eigenvector. Black ellipses indicate the avoided-crossing behavior.

Interestingly, an anomaly in typical 2D materials featured mode, flexural acoustic phonon (ZA) [37], is observed. ZA shows a linear dispersive curve along Γ–X, while it is quadratic along Γ–Y. Such behavior is different from that in other equilateral 2D-layered structures (e.g., equilateral pentagonal and hexagonal structures) and monolayered structures, where the ZA mode shows a quadratic-like curve in the basal plane [8, 9, 12, 13, 27]. Here, we attribute this dispersive difference to the bucking Se atoms in the zigzag- and flat-type chains, forming non-planar structures in the basal plane, leading to the non-pure out-of-plane vibration of the atoms (deviate from $c$-axis) near the zone center ($\boldsymbol{q} \rightarrow 0$) [38, 39] (Figure S3 and Table SI). Thus, the ZA branches are not purely quadratic near Γ-point and possess nonzero group velocities [40]. The transverse acoustic (TA) branch goes steeply up to 9 meV along Γ–Y. However, it is bent by the transverse optical mode vibrating along out-of-plane (TO$_{z1}$) and ends only at 3 meV at X-point, known as the avoided-crossing behavior (black ellipses Figure 2d) [41, 42]. The longitudinal acoustic (LA) branch exhibits the avoided-crossing behavior with TO$_{z3}$ along Γ–X and with TO$_{z1}$ along Γ–Y. The 4$^{th}$ mode, behaving as the TO$_y$ mode along Γ–X and the longitudinal optical (LO) mode along Γ–Y, shows the different slopes along the high-symmetry directions. In addition, the 5$^{th}$ mode, identified as the LO mode along Γ–X and TO$_x$ mode along Γ–Y, also exhibits a large difference of slopes along the high-symmetry directions in the calculation. Figure 2d provides a clear dispersive behavior of each phonon mode sorted by their eigenvectors, from which the detailed in-plane dispersion difference can be observed obviously.

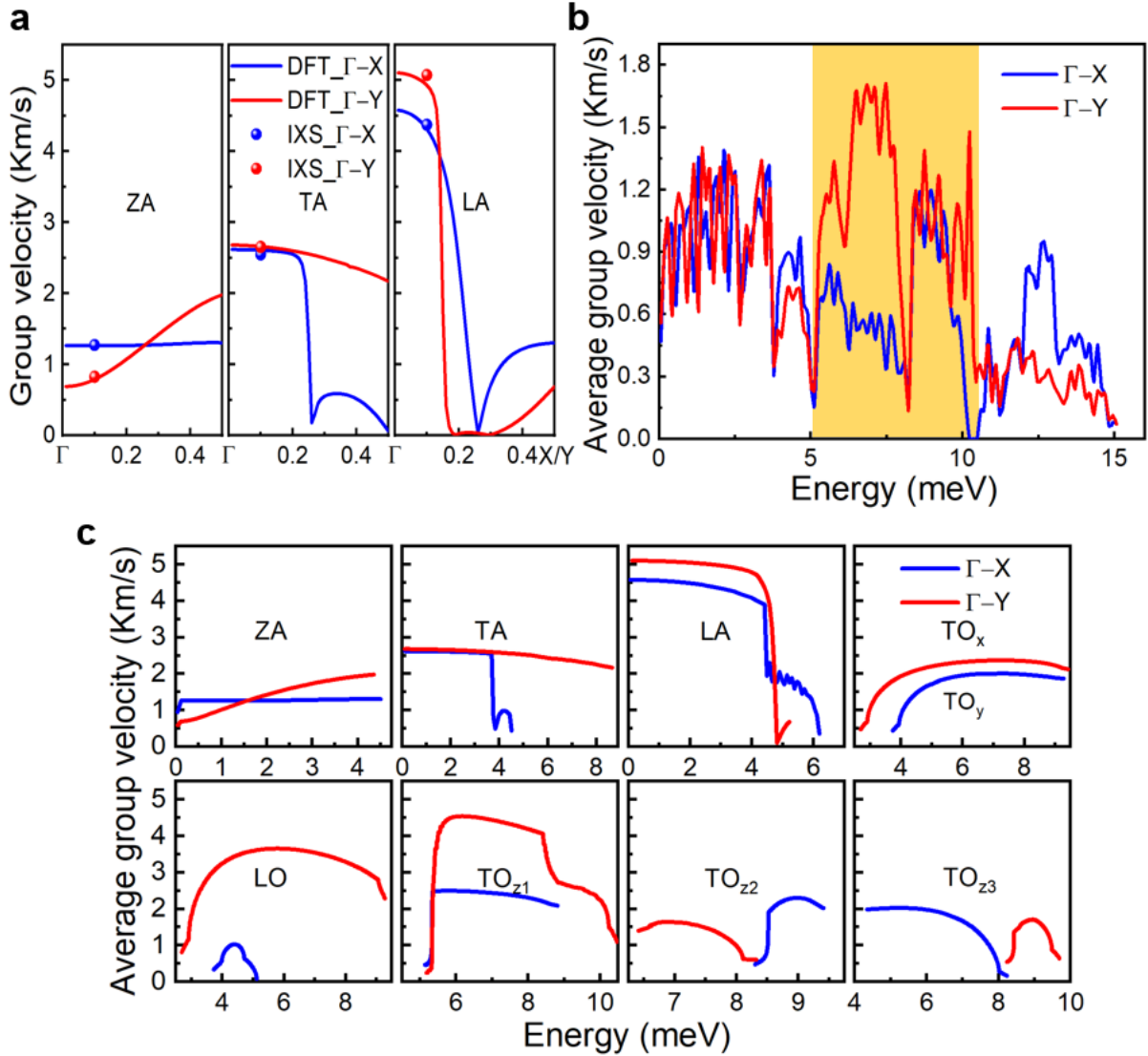

Figure 3. Phonon group velocities of PdSe$_2$ along Γ–X and Γ–Y. **a** *q*-dependent absolute values of phonon group velocities of the acoustic phonons along Γ–X and Γ–Y. Balls are the velocities extracted from IXS at *q* = 0.1 (r.l.u.). **b** Energy-dependent average group velocity of phonon modes along Γ–X and Γ–Y, respectively, showing high anisotropy for phonons between 5 and 10 meV. Yellow shadow is the guide for eyes. **c** Energy-dependent average group velocity of each LE mode along Γ–X and Γ–Y.

The significant in-plane phonon dispersion differences determine the high anisotropy in phonon group velocities. As shown in Figure 3a, the *q*-dependent group velocity of ZA mode ($v_{ZA}$, similar definitions for other group velocities of each mode) shows a near-constant value along Γ–X, while it monotonically increases along Γ–Y. It should be noted that ZA mode has higher group velocity along Γ–X than along Γ–Y near zone center due to its linear dispersion behavior, as confirmed by the extracted velocities from IXS measurements. $v_{TA}$ exhibits a knee point along Γ–X due to the avoided-crossing behavior and thus has a much lower value there. In order to evaluate the anisotropy of the group velocities, the energy-dependent average group velocity is calculated as [43]:

$$v(\omega) = \frac{(v^\mu)^2}{|v|}(\omega) = \left(\frac{\Delta q}{2\pi}\right)^3 \sum_\lambda \frac{(v_\lambda^\mu)^2}{|v_\lambda^\mu|} \delta(\omega - \omega_\lambda) / \sum_j g_j(\omega) \tag{1}$$

where $v_\lambda^\mu$ is the group velocity of phonon mode $\lambda$ at wavevector $q$ along the $\mu$-axis and $g_j(\omega)$ is the phonon density of state of mode $j$. Surprisingly, the group velocities along Γ–Y are significantly higher than that along Γ–X in the energy range from 5 to 11 meV (Figure 3b), suggesting a large phonon anisotropy in the basal plane. Such velocity anisotropy mainly comes from the unique dispersion of TA, LO, TO$_{y/x}$, and TO$_{z1}$ modes (Figure 3c).

Based on the equation of the lattice thermal conductivity tensor ($\kappa$), phonon group velocity dominates the anisotropic thermal transport, expressed as[44]

$$\kappa_{\alpha\beta} = \frac{1}{3} \sum_\lambda c_\lambda v_\lambda^\alpha v_\lambda^\beta \tau_\lambda \tag{2}$$

where $c_\lambda$ is the specific heat capacity of phonon mode $\lambda$, $v_\lambda^{\alpha(\beta)}$ is the group velocity of phonon mode $\lambda$ along the $\alpha(\beta)$ direction, and $\tau_\lambda$ is relaxation time of phonon mode $\lambda$. Thus, there is no doubt that such high anisotropy in phonon group velocity will lead to a high anisotropy in IPTC. As expected, it can be found that $\kappa_b$ is much higher than $\kappa_a$ from the calculated temperature-dependent anisotropic lattice thermal conductivity in Figure 4a. The anisotropy ratio, $\kappa_b/\kappa_a$, is almost constant at 1.8 over the temperature range of 300 ~ 800 K.

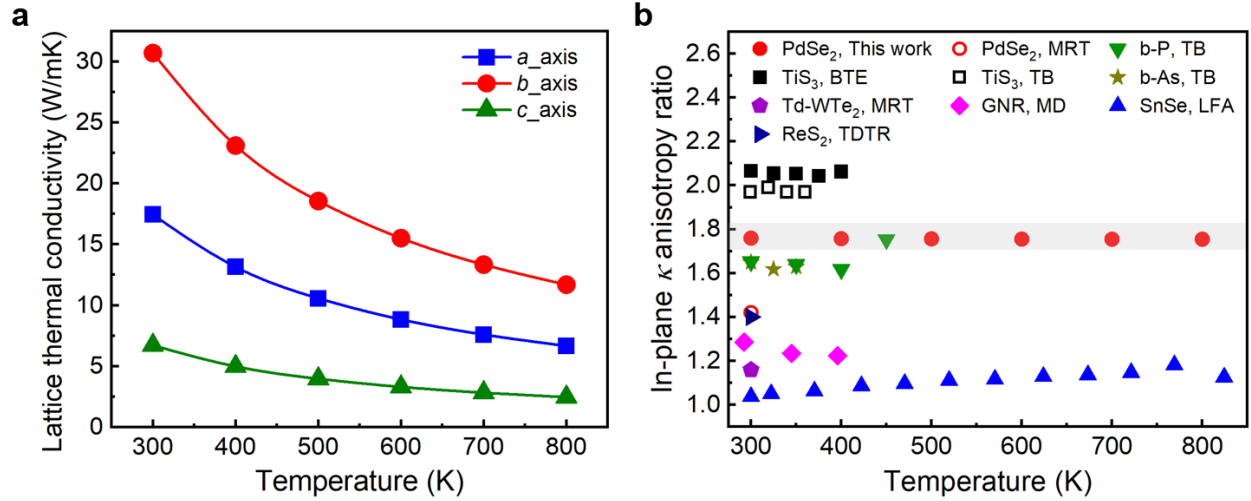

Figure 4. Anisotropic thermal conductivity of PdSe$_2$. **a** Temperature-dependence of the anisotropic lattice thermal conductivity from the BTE calculations. **b** In-plane thermal conductivity anisotropy ratio of various 2D-layered materials versus temperature. The samples used for measurements are 1.62 nm thickness for PdSe$_2$, 150 nm for ReS$_2$, 11.2 nm for Td-WTe$_2$, 100 nm for TiS$_3$, 31 nm for b-P, bulk for SnSe, and 124 nm thick nanoribbon for b-As. Abbreviations: Thermal bridge (TB), graphene nanoribbon (GNR), molecular dynamics (MD), laser flash apparatus (LFA), and time-domain thermoreflectance (TDTR). Grey shadow indicates the result of this work. Other data shown here are from refs. [9, 24, 45-47].

As shown in Figure 4b, this anisotropy ratio is quite large and comparable to some of the values in most in-plane anisotropic layered materials, such as 1.6 for black phosphorus [9, 45] 1.4 for ReS$_2$ [46], 1.3 for graphene nanoribbon [47], and comparable to the highest score of 2.0 for TiS$_3$ up to date [11]. As introduced above, the in-plane lattice constants, Pd-Se bond lengths, bond strengths, and the bond force constants are nearly equal in this nearly-equilateral in-plane structured PdSe$_2$. It is neither the cases in ReS$_2$ and TiS$_3$, where the difference between the weak bonding and the covalent bonding directions induces the IPTC anisotropy, and nor the case in the equilateral graphene, where the IPTC anisotropy comes from the difference between the zigzag- and armchair-directions rather than the crystallographic axis directions. Table I lists the anisotropy of IPTC in the materials shown in Figure 4b, which shows the unique anisotropy in IPTC of PdSe$_2$, i.e., high IPTC anisotropy with nearly-equal in-plane lattice constants and the nearly-isotropic bonding.

Table I. Anisotropic information of various 2D layered materials summarized from refs. [9, 24, 45-47]. The lattice ratios of GNR, b-P, and b-As are defined by the structure geometry.

| Materials | In-plane directions | | Lattice ratios | Bonding | IPTC Anisotropy ratios |
|---|---|---|---|---|---|
| PdSe$_2$ | $a$-axis | $b$-axis | 0.98 | nearly-isotropic | 1.79 this work; 1.42 Exp. |
| TiS$_3$ | $a$-axis | $b$-axis | 0.69 | anisotropic | 2.04 Exp. |
| ReS$_2$ | $a$-axis | $b$-axis | 0.98 | anisotropic | 1.40 Exp. |
| Td-WTe | $a$-axis | $b$-axis | 0.56 | anisotropic | 1.16 Exp. |
| SnSe | $c$-axis | $b$-axis | 0.93 | anisotropic | 1.04 Exp. |
| GNR | zigzag | armchair | 0.87 | isotropic | 1.29 Cal. |
| b-P | zigzag | armchair | 0.87 | isotropic | 1.65 Exp. |
| b-As | zigzag | armchair | 0.87 | isotropic | 1.65 Exp. |

It should be mentioned that our calculation shows a moderate difference with the measurement in PdSe$_2$, where an anisotropy ratio of IPTC up to 1.42 was reported [24]. There could be two reasons that caused this discrepancy: (1) The bulk crystal model instead of the multilayer was used here for the calculation to adapt to our IXS experiment because phonon dispersions play a crucial role in computing the correct thermal conductivity [50]. Thus, other phonon scattering mechanisms, including phonon-boundary scattering and phonon-defect scattering, were ignored in our calculations. (2) The high electrical conductivity [51] will increase the thermal conductivity of PdSe$_2$. The electronic thermal conductivity was not extracted from the measurement in ref. [24], making a relatively high impact on the result of the anisotropy of IPTC. (3) The nonequilibrium phonons and temperature and strain effects are unavoidable to measure the IPTC using micro-Raman thermometry, sometimes leading to inaccurate results.

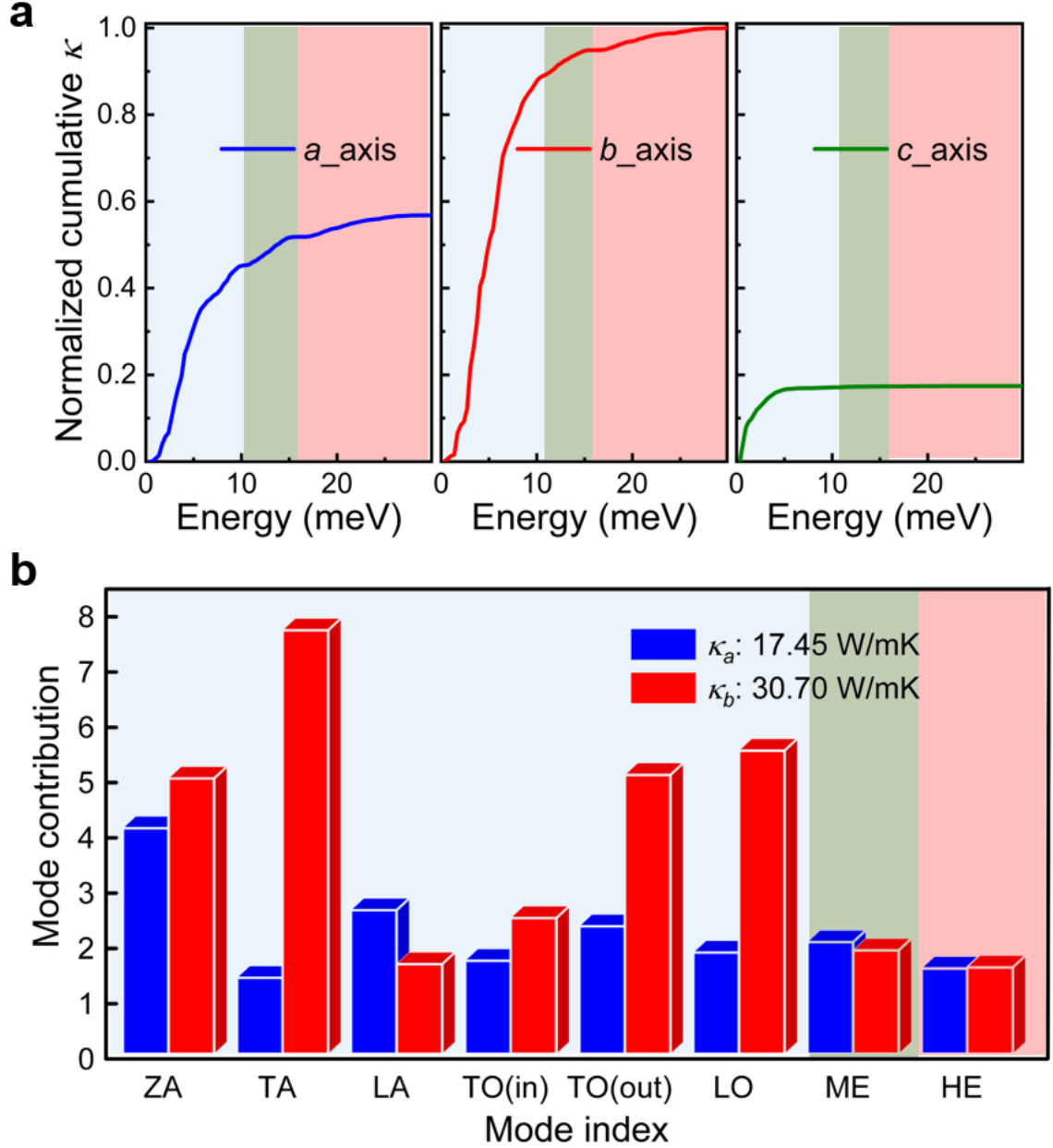

Figure 5. Contributions on the anisotropic lattice thermal conductivity of PdSe$_2$. **a** Phonon energy-dependence of the anisotropic cumulative lattice thermal conductivity at 300 K. **b** Mode contribution to thermal conductivity along $a$- and $b$-axis. Azure, dark-sea-green, and light-pink shadows represent the LE -, ME -, and HE-phonons, respectively.

In order to explore the anisotropy in IPTC, the energy-dependent anisotropic cumulative lattice thermal conductivities of PdSe$_2$ at 300 K are shown in Figure 5a. It is found that along $a$- and $b$-axis, the LE phonons below 10 meV dominate the thermal conductivities by accounting for 90% of the contribution, while the middle-energy (ME, 11–16 meV) and the high-energy (HE, 16–32 meV) optical phonons are responsible for the rest. The differences in the value and trend of cumulative lattice thermal conductivities suggest the significant anisotropy in LE phonon propagations along Γ–X and Γ–Y, consistent with the phonon dispersion and group velocity anisotropies discussed above. It should be noted that $\kappa_c$ has the lowest value and cut-off phonon energy in the cumulative thermal conductivity due to the weak interlayer vdW interactions. To further distinguish the role of phonons in the anisotropic IPTC, we calculated the mode contributions on the thermal conductivity, as shown in Figure 5b and Figure S4b. It is found that the mode contributions have a significant difference in the two high-symmetry directions. Interestingly, TA, TO$_{z1-z3}$, and LO branches have significantly greater values along $b$-axis, and possess the anisotropy ratios of 5.57, 2.19, and 2.99, respectively, making them the main contribution on the anisotropic IPTC in PdSe$_2$ (Figure S4). ZA and TO$_{y/x}$ branches have slightly higher values along $b$-axis with the anisotropic ratios of 1.22 and 1.46. In addition, LA branch has a slightly lower value along $b$-axis than along $a$-axis with an anisotropic ratio of 0.63, while the ME- and HE-optical branches have comparable values along the two high-symmetry directions. These significant anisotropic contributions can be well revealed by the different dispersion behaviors and group velocities discussed above.

The effect of the phonon scattering rate should also be taken into account, although the group velocity can successfully explain the high anisotropy in IPTC. Figure 6 shows the extracted phonon linewidths (scattering rates) from IXS data after deconvoluting with the instrument resolution by Gaussian fitting. It can be seen that the calculated anharmonic scattering rates of the LE-phonons are slightly smaller than those from experiments due to the neglect of other scattering processes in calculations, such as the phonon-defect scattering and phonon-boundary scattering. Although the anisotropy of scattering rates is not as remarkable as that of group velocities, there still exists differences for the acoustic modes. For ZA along Γ–Y, fewer scattering channels are observed due to the quadratic dispersive behavior and thus show low scattering rates [52]. The higher scattering rates emerge for TA and LA mainly caused by the avoided-crossing behavior, which flattens the acoustic branch at high $q$. The strong scattering rates concentrate in the HE-phonons energy region but contribute little to thermal conductivity (Figure S5). The scattering channels show that the three-phonon scattering of LE-phonons is limited to some extent (inset in Figure S5). The frozen phonon potentials show that from Γ-point to the zone boundary, most of the LE-phonons behave the near-quadratic properties (Figure S6), indicating weak anharmonicity of the LE-phonons. Therefore, phonon scattering rates have little impact on the IPTC anisotropy.

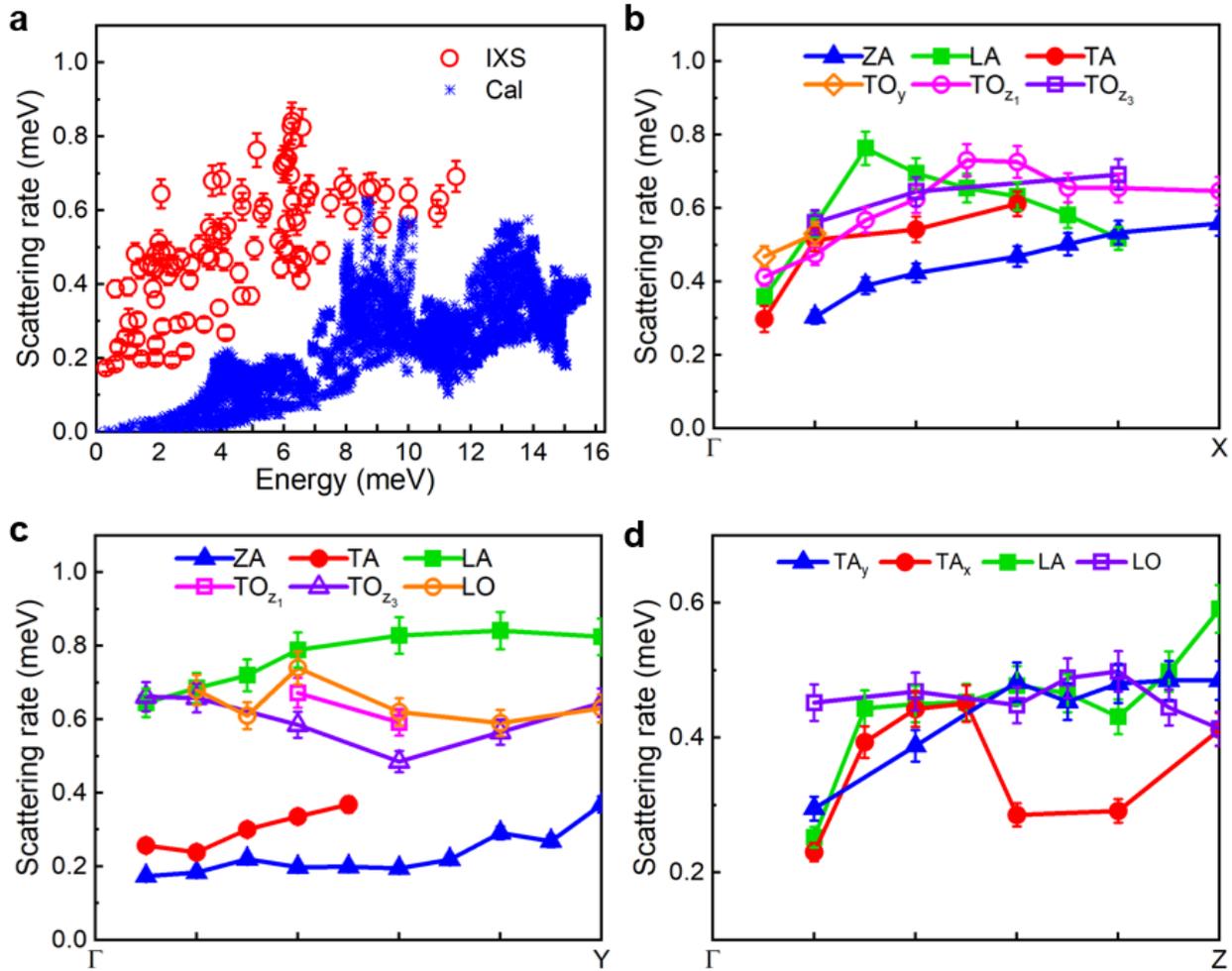

Figure 6. Phonon scattering rates in layered $PdSe_2$. **a** Comparison between the experimental and calculated energy-dependent scattering rates. **b**, **c**, and **d** are the $q$-dependent scattering rates extracted from IXS data along Γ–X, Γ–Y, and Γ–Z, respectively. Error bars are the fitting uncertainty.

Due to the nearly-equal in-plane lattice constants and the nearly-isotropic bonding of PdSe$_2$, it is difficult to investigate the origin of the IPTC anisotropy through the typical crystal structure difference [9, 45-47]. As reported, the bucking structure could significantly change the dispersive behavior in the phonon spectrum [38-40]. In PdSe$_2$, the bucking structure induces the unique vibration of the LE-phonons with the comparable displacements of the heavier Pd and lighter Se atoms (Figures S7-S10), leading to the large anisotropy of dispersions along Γ–X and Γ–Y. For example, in the LO mode, it is found that the atoms vibrate nearly along the zigzag edge directions along $a$-axis, especially at high $q$-points, leading to a "bending-like" dispersive behavior along Γ–X. However, they vibrate nearly along $b$-axis, leading to a typical dispersive mode along Γ–Y, and thus the anisotropic phonon dispersion behavior. Detailed descriptions of other modes can be found in the SI. It is complex to distinguish the bucking effect on each phonon mode exhibiting the anisotropy (e.g., ZA, TA, TO, LO, and TO$_{z1}$).

Table II. Mode force constants (meV/Å$^2$) of the LE-optical mode at $q$ = 0, 0.1, and 0.3 along the two high-symmetry directions, obtained by the second derivation of the frozen phonon potentials. Pd$_i$ ($i = a$, $b$, $c$) shown here as the reference of the atomic displacement in each mode.

| Mode<br>$q$ | ZA | | TO | | LO | | TO$_{z1}$ | | TO$_{z2}$ | | TO$_{z3}$ | |
|---|---|---|---|---|---|---|---|---|---|---|---|---|
| | Γ–X | Γ–Y | Γ–X | Γ–Y | Γ–X | Γ–Y | Γ–X | Γ–Y | Γ–X | Γ–Y | Γ–X | Γ–Y |
| Γ-point | Pd$_c$ | Pd$_c$ | Pd$_b$ | Pd$_a$ | Pd$_a$ | Pd$_b$ | Pd$_c$ | Pd$_c$ | Pd$_c$ | Pd$_c$ | Pd$_c$ | Pd$_c$ |
| | — | — | 170 | 320 | 320 | 170 | 260 | 260 | 310 | 310 | 310 | 310 |
| 0.1 | Pd$_c$ | Pd$_c$ | Pd$_b$ | Pd$_a$ | Pd$_a$ | Pd$_b$ | Pd$_c$ | Pd$_c$ | Pd$_c$ | Pd$_c$ | Pd$_c$ | Pd$_c$ |
| | 20 | 10 | 190 | 220 | 200 | 220 | 260 | 270 | 340 | 350 | 350 | 360 |
| 0.3 | Pd$_c$ | Pd$_c$ | Pd$_b$ | Pd$_c$ | Pd$_c$ | Pd$_b$ | Pd$_b$ | Pd$_a$ | Pd$_{ac}$ | Pd$_a$ | Pd$_{ac}$ | Pd$_c$ |
| | 40 | 20 | 260 | 430 | 320 | 410 | 270 | 370 | 540 | 860 | 680 | 1020 |

Here, the mode force constant (MFC) is used to understand this anisotropy, which is obtained by extracting the second derivative of frozen phonon potentials. MFC reflects all the atomic vibration information, including the effect of the bucking atoms in a particular mode for the selected momentum, and thus can evaluate the strength of the vibration of the phonon mode. Table II lists the MFCs of the LE-optical modes at $q$ = 0, 0.1, and 0.3 along both the two high-symmetry directions. The MFCs are obtained based on the heavier Pd atom as the reference. It is found that for the ZA mode, the MFCs of Pd$c$ increase with $q$ and show about two times along Γ–X than along Γ–Y. For other modes, the vibration direction of Pd atoms changes with $q$ and show higher values along Γ–Y than along Γ–X. For example, the MFCs of Pd atoms in TO, LO, TO$_{z1}$, TO$_{z2}$, and TO$_{z3}$ modes at q = 0.3 show 1.7, 1.3, 1.4, 1.6, and 1.5 times along Γ–Y than those along Γ–X, which are responsible for the difference in group velocities. In addition, the avoided-crossing behaviors caused by the optical-acoustic phonon eigenvector hybridization [53], are also significant for lowering the group velocities [54,55] and changing the dispersion configurations. The different avoided-crossing behaviors in dispersions are mainly attributed to the bucking structures and the structure geometry [56], which needs further investigation in PdSe$_2$ in the future.

## IV. CONCLUSION

In summary, a giant IPTC anisotropy with a ratio of about 1.8 is discovered in the pentagonal structure PdSe$_2$. Unlike most layered materials where the IPTC anisotropy originates from anisotropic structures, our findings show that the IPTC anisotropy in PdSe$_2$ emerges in the nearly-equilateral in-plane structure with nearly-isotropic bonding. The IXS measurements show a significant difference in the LE-phonon dispersions and the group velocities between along Γ–X and Γ–Y directions), further revealed by the first-principles calculations. The lower group velocities and the avoided-crossing behaviors along Γ–X induce lower lattice thermal conductivity along *a*-axis, closely related to the bucking structure in the basal plane. Our results reveal a new viewpoint for anisotropic IPTC in the nearly-equilateral structures with nearly-isotropic bonding in 2D-layered materials, pave a new path to find materials with anisotropic thermal conductivity, and provide a significant reference for thermal management in electronic devices.

**Conflict of interest**

The authors declare no conflict of interest.

**Acknowledgements**

The work at Beijing Institute of Technology is supported by the National Natural Science Foundation of China with Grant No. 11572040 and Beijing Natural Science Foundation (Grant No. Z190011). The work at Henan Polytechnic University is supported by the Doctoral Foundation of Henan Polytechnic University (Natural Science). Chen Li thanks the Initial Complement of University of California, Riverside. This research used the resource of the Advanced Photon Source, a US Department of Energy (DOE) Office of Science User Facility, operated for the DOE Office of Science by Argonne National Laboratory under Contract No. DE-AC02-06CH11357.

**Author Contributions**

BW conceived this project. BW and JH designed the experiments. BW, QC, and AHS performed the inelastic X-ray scattering measurements. BW, JL performed the theoretical calculation. BW, QC, JL, JH, and CL discussed the data. BW wrote the manuscript. All authors contributed to discussing the data and editing the manuscript.

**Reference:**

[1] A. H. Castro Neto, F. Guinea, N. M. R. Peres, K. S. Novoselov and A. K. Geim, Rev. Mod. Phys. **81**, 109-162 (2009).
[2] X. Wang, Z. Song, W. Wen, H. Liu, J. Wu, C. Dang, M. Hossain, M. A. Iqbal and L. Xie, Adv. Mater. **31** (2018).
[3] K. Novoselov, o. A. Mishchenko, o. A. Carvalho and A. C. Neto, Science 353 (2016).
[4] C. Kim, J. C. Park, S. Y. Choi, Y. Kim, S. Y. Seo, T. E. Park, S. H. Kwon, B. Cho and J. H. Ahn, Small **14**, e1704116 (2018).
[5] S. Goossens, G. Navickaite, C. Monasterio, S. Gupta, J. J. Piqueras, R. Pérez, G. Burwell, I. Nikitskiy, T. Lasanta, T. Galán, E. Puma, A. Centeno, A. Pesquera, A. Zurutuza, G. Konstantatos and F. Koppens, Nature Photonics 11, 366-371 (2017).
[6] D. Pesin and A. H. MacDonald, Nat Mater **11**, 409-16 (2012).
[7] W. Han, R. K. Kawakami, M. Gmitra and J. Fabian, Nat Nanotechnol **9**, 794-807 (2014).
[8] L. Lindsay, D. Broido and N. Mingo, Phys. Rev. B **82**, 115427 (2010).
[9] S. Lee, F. Yang, J. Suh, S. Yang, Y. Lee, G. Li, H. Sung Choe, A. Suslu, Y. Chen, C. Ko, J. Park, K. Liu, J. Li, K. Hippalgaonkar, J. J. Urban, S. Tongay and J. Wu, Nat. Commun. **6**, 8573 (2015).
[10] C. W. Li, J. Hong, A. F. May, D. Bansal, S. Chi, T. Hong, G. Ehlers and O. Delaire, Nat. Phys. **11**, 1063-1069 (2015).
[11] H. Liu, X. Yu, K. Wu, Y. Gao, S. Tongay, A. Javey, L. Chen, J. Hong and J. Wu, Nano Lett. **20**, 5221-5227 (2020).
[12] Zhang, S., Zhou, J., Wang, Q., Chen, X., Kawazoe, Y., & Jena, P. (2015). Penta-graphene: A new carbon allotrope. Proceedings of the National Academy of Sciences, 112(8), 2372-2377.
[13] Aierken, Y., Leenaerts, O., & Peeters, F. M. (2016). A first-principles study of stable few-layer penta-silicene. Physical Chemistry Chemical Physics, 18(27), 18486-18492.
[14] Deng, S., Li, L., & Zhang, Y. (2018). Strain modulated electronic, mechanical, and optical properties of the monolayer PdS2, PdSe2, and PtSe2 for tunable devices. ACS Applied Nano Materials, 1(4), 1932-1939.
[15] Deng, S., Li, L., & Zhang, Y. (2018). Strain modulated electronic, mechanical, and optical properties of the monolayer PdS2, PdSe2, and PtSe2 for tunable devices. ACS Applied Nano Materials, 1(4), 1932-1939.
[16] M. Sun, J.-P. Chou, L. Shi, J. Gao, A. Hu, W. Tang and G. Zhang, ACS Omega **3**, 5971-5979 (2018).
[17] W. Lei, S. Zhang, G. Heymann, X. Tang, J. Wen, X. Zheng, G. Hu and X. Ming, Journal of Materials Chemistry C **7**, 2096-2105 (2019).
[18] Y.-S. Lan, X.-R. Chen, C.-E. Hu, Y. Cheng and Q.-F. Chen, Journal of Materials Chemistry A **7**, 11134-11142 (2019).
[19] Lu, L. S., Chen, G. H., Cheng, H. Y., Chuu, C. P., Lu, K. C., Chen, C. H., ... & Chang, W. H. (2020). Layer-Dependent and in-Plane Anisotropic Properties of Low-Temperature Synthesized Few-Layer PdSe2 Single Crystals. ACS nano, 14(4), 4963-4972.
[20] Q. Liang, Q. Wang, Q. Zhang, J. Wei, S. X. Lim, R. Zhu, J. Hu, W. Wei, C. Lee, C. Sow, W. Zhang and A. T. S. Wee, Adv. Mater. **31**, 1807609 (2019).
[21] D. Qin, P. Yan, G. Q. Ding, X. J. Ge, H. Y. Song and G. Y. Gao, Scientific Reports **8**, 8, 2764 (2018).
[22] A. A. Puretzky, A. D. Oyedele, K. Xiao, A. V. Haglund, B. G. Sumpter, D. Mandrus, D. B. Geohegan and L. Liang, 2D Materials **5** (2018).
[23] W. Luo, A. D. Oyedele, Y. Gu, T. Li, X. Wang, A. V. Haglund, D. Mandrus, A. A. Puretzky, K. Xiao, L. Liang and X. Ling, Adv. Funct. Mater. **30**, 2003215 (2020).
[24] L. Chen, W. Zhang, H. Zhang, J. Chen, C. Tan, S. Yin, G. Li, Y. Zhang, P. Gong, and L. Li. Sustainability 2021, 13, 4155.
[25] C. Soulard, X. Rocquefelte, P. E. Petit, M. Evain, S. Jobic, J. P. Itié, P. Munsch, H. J. Koo and M. H. Whangbo, lnorg. Chem. **43**, 1943-1949 (2004).
[26] Lv, P., Tang, G., Liu, Y., Lun, Y., Wang, X., & Hong, J. (2021). Van der Waals direction transformation induced by shear strain in layered PdSe2. Extreme Mechanics Letters, 101231.
[27] Liu, H., Qin, G., Lin, Y., & Hu, M. (2016). Disparate strain dependent thermal conductivity of two-dimensional penta-structures. Nano letters, 16(6), 3831-3842.
[28] Koskinen, Pekka, Sami Malola, and Hannu Häkkinen. "Evidence for graphene edges beyond zigzag and armchair." Physical Review B 80.7 (2009): 073401.

[29] Jin, Y.; Li, X.; Yang, J. Single layer of MX3 (M = Ti, Zr; X = S, Se, Te): a new platform for nano-electronics and optics. Phys. Chem. Chem. Phys. 2015, 17, 18665−18669.
[30] Toellner, T., Alatas, A. & Said, A. Six‐reflection meV‐monochromator for synchrotron radiation. J Synchrotron Radiat 18, 605-611 (2011).
[31] Said, A. H., Sinn, H. & Divan, R. New developments in fabrication of high-energy resolution analyzers for inelastic X-ray spectroscopy. J Synchrotron Radiat 18, 492-496 (2011).
[32] P. E. Blöchl, Phys. Rev. B 50, 17953 (1994).
[33] Kresse, G. & Furthmüller, J. Efficiency of ab-initio total energy calculations for metals and semiconductors using a plane-wave basis set. Phys. Rev. B 54, 11169 (1996); Comput. Mater. Sci. 6, 15-50 (1996).
[34] J. Klimeš, D. R. Bowler, and A. Michaelides, Chemical accuracy for the van der waals density functional, J. Phys.: Condens. Matter 22, 022201 (2009).
[35] Togo, A., Oba, F. & Tanaka, I. First-principles calculations of the ferroelastic transition between rutile-type and -type at high pressures. Phys. Rev. B 78, 134106 (2008).
[36] W. Li, J. Carrete, N. A. Katcho, and N. Mingo, Comput. Phys. Commun. 185, 1747 (2014).
[37] Lindsay, L., D. A. Broido, and Natalio Mingo. "Flexural phonons and thermal transport in graphene." Physical Review B 82.11 (2010): 115427.
[38] Xie, H., Ouyang, T., Germaneau, É., Qin, G., Hu, M., & Bao, H. (2016). Large tunability of lattice thermal conductivity of monolayer silicene via mechanical strain. Physical Review B, 93(7), 075404.
[39] Zhang, X., Xie, H., Hu, M., Bao, H., Yue, S., Qin, G., & Su, G. (2014). Thermal conductivity of silicene calculated using an optimized Stillinger-Weber potential. Physical Review B, 89(5), 054310.
[40] X. Gu and R. Yang, J. Appl. Phys. 117, 025102 (2015)
[41] Christensen, M., Abrahamsen, A. B., Christensen, N. B., Juranyi, F., Andersen, N. H., Lefmann, K., ... & Iversen, B. B. (2008). Avoided crossing of rattler modes in thermoelectric materials. Nature materials, 7(10), 811-815.
[42] Zhu, Y., Wei, B., Liu, J., Koocher, N. Z., Li, Y., Hu, L., ... & Hong, J. (2021). Physical insights on the low lattice thermal conductivity of AgInSe2. Materials Today Physics, 19, 100428.
[43] Mukhopadhyay, S.; Lindsay, L.; Singh, D. J. Optic phonons and anisotropic thermal conductivity in hexagonal Ge2Sb2Te5. Sci. Rep. 2016, 6, 37076.
[44] Broido, D. A.; Malorny, M.; Birner, G.; Mingo, N.; Stewart, D. A. Intrinsic lattice thermal conductivity of semiconductors from first principles. Appl. Phys. Lett. 2007, 91, 231922.
[45] Jeon, S. G., Shin, H., Jaung, Y. H., Ahn, J., & Song, J. Y. (2018). Thickness-dependent and anisotropic thermal conductivity of black phosphorus nanosheets. Nanoscale, 10(13), 5985-5989.
[46] Jang, H.; Ryder, C.R.; Wood, J.D.; Hersam, M.C.; Cahill, D.G. 3D Anisotropic Thermal Conductivity of Exfoliated Rhenium Disulfide. Adv. Mater. 2017, 29, 1700650.
[47] Hu, J.; Ruan, X.; Chen, Y. P. Thermal conductivity and thermal rectification in graphene nanoribbons: a molecular dynamics study. Nano Lett. 2009, 9, 2730−2735.
[48] Chen, Y.; Chen, C.; Kealhofer, R.; Liu, H.; Yuan, Z.; Jiang, L.; Suh, J.; Park, J.; Ko, C.; Choe, H. S.; Avila, J.; Zhong, M.; Wei, Z.; Li, J.; Li, S.; Gao, H.; Liu, Y.; Analytis, J.; Xia, Q.; Asensio, M. C.; Wu, J. Black arsenic: a layered semiconductor with extreme in-plane anisotropy. Adv. Mater. 2018, 30, 1800754.
[49] Zhao, L.-D.; Lo, S.-H.; Zhang, Y.; Sun, H.; Tan, G.; Uher, C.; Wolverton, C.; Dravid, V. P.; Kanatzidis, M. G. Ultralow thermal conductivity and high thermoelectric figure of merit in SnSe crystals. Nature 2014, 508, 373−377.
[50] L. Paulatto, F. Mauri, and M. Lazzeri, Phys. Rev. B 87, 214303 (2013).
[51] Zhao, Y., Yu, P., Zhang, G., Sun, M., Chi, D., Hippalgaonkar, K., ... & Wu, J. (2020). Low‐Symmetry PdSe2 for High Performance Thermoelectric Applications. Advanced Functional Materials, 30(52), 2004896.
[52] Lindsay L, Li W, Carrete J, et al. Phonon thermal transport in strained and unstrained graphene from first principles[J]. Physical Review B, 2014, 89: 155426.
[53] Li, W., Carrete, J., Madsen, G. K., & Mingo, N. (2016). Influence of the optical-acoustic phonon hybridization on phonon scattering and thermal conductivity. Physical Review B, 93(20), 205203.
[54] Christensen, M., Abrahamsen, A. B., Christensen, N. B., Juranyi, F., Andersen, N. H., Lefmann, K., ... & Iversen, B. B. (2008). Avoided crossing of rattler modes in thermoelectric materials. Nature materials, 7(10), 811-815.
[55] Zhu, Y., Wei, B., Liu, J., Koocher, N. Z., Li, Y., Hu, L., ... & Hong, J. (2021). Physical insights on the low lattice thermal conductivity of AgInSe2. Materials Today Physics, 19, 100428.
[56] Dove, M. T., & Dove, M. T. (1993). Introduction to lattice dynamics (No. 4). Cambridge university press.

# Supporting information
# Giant Anisotropic in-Plane Thermal Conduction Induced by Anomalous Phonons in Nearly-Equilaterally Structured PdSe$_2$

Bin Wei[1,2*], Junyan Liu[2*], Qingan Cai[3], Ahmet Alatasaid[4], Chen Li[3,5†], Jiawang Hong[2†]

[1]Henan Key Laboratory of Materials on Deep-Earth Engineering, School of Materials Science and Engineering, Henan Polytechnic University, Jiaozuo 454000, China

[2]School of Aerospace Engineering, Beijing Institute of Technology, Beijing 100081, China

[3]Mechanical Engineering, University of California, Riverside, Riverside, CA 92521, USA

[4]Advanced Photon Source, Argonne National Laboratory, Argonne, IL 60439, USA

[5]Materials Science and Engineering, University of California, Riverside, Riverside, CA 92521, USA

[*]The authors contribute equally
[†]Corresponding author: chenli@ucr.edu.cn; hongjw@bit.edu.cn

Figure S1 shows the sample information of PdSe$_2$ for IXS measurements. The narrow full-width-at-half-maximum (FWHM) indicates the high quality of the crystal.

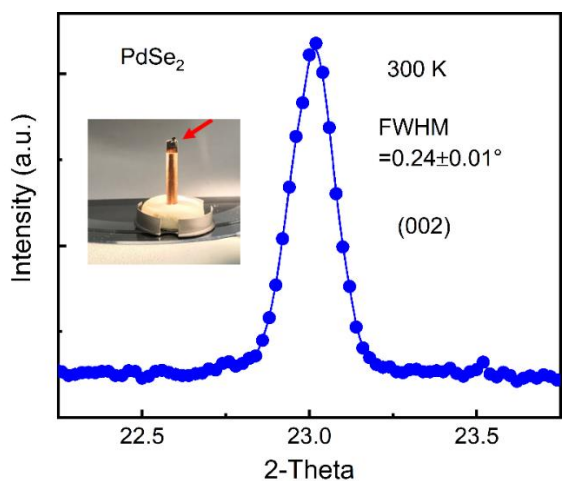

Figure S1. Rocking curve of PdSe$_2$ single crystal for Bragg peak at (002) showing the high quality of the sample. Inset is the exploited sample with 60 μm thickness attached on a copper post for IXS measurement.

Figure S2a shows the calculated phonon dispersions of PdSe$_2$ throughout the Brillouin zone (Figure S2b), which is sorted by three groups based on the phonon energy, i.e., the low-energy (LE) phonons (including the acoustic and LE-optical phonons) with dispersive branches, the middle-energy (ME) and the high-energy (HE) phonons with non-dispersive branches.

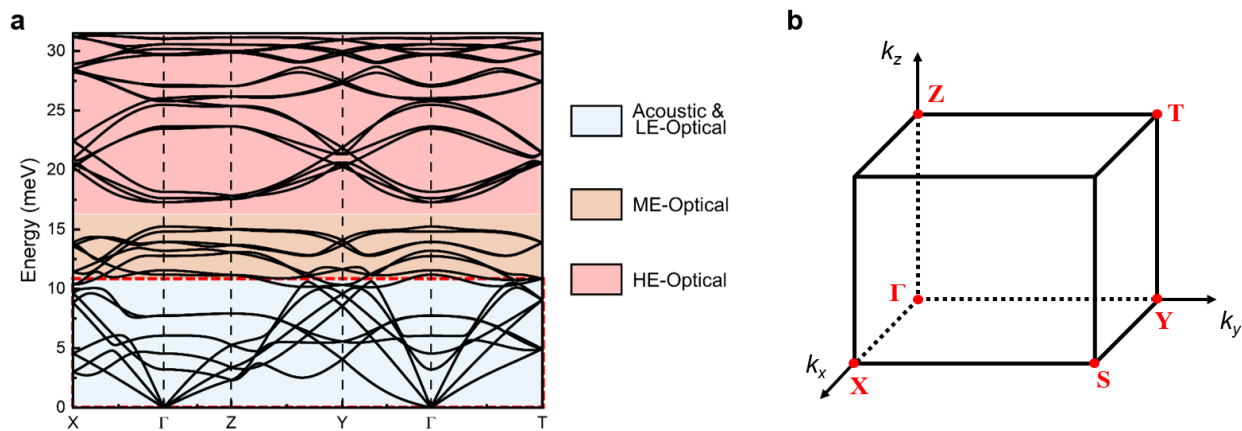

Figure S2. Entire phonon dispersion (a) and the Brillouin zone (b) of PdSe$_2$. Azure, bisque, and light-pink shadows represent the acoustic and LE-optical, ME-optical, and HE-optical phonons. The dashed red box represents the LE phonon dispersion of PdSe$_2$.

Figure S3a shows the degree of the derivation of the real part in ZA mode eigenvector heavier Pd atom. The larger derivation along Γ–X indicates the linear-like dispersion of ZA, while the little derivation along Γ–Y indicates a non-pure quadratic dispersion of ZA. Figure S3b and S3c show the vibration pattern of ZA along *a*- (upper) and *b*-axis (bottom) at $q = 0.1$ and 0.4, respectively. It can be seen that Pd and Se atoms have similar displacements, and the derivation of the atoms from *c*-axis is larger along *a*-axis than *b*-axis.

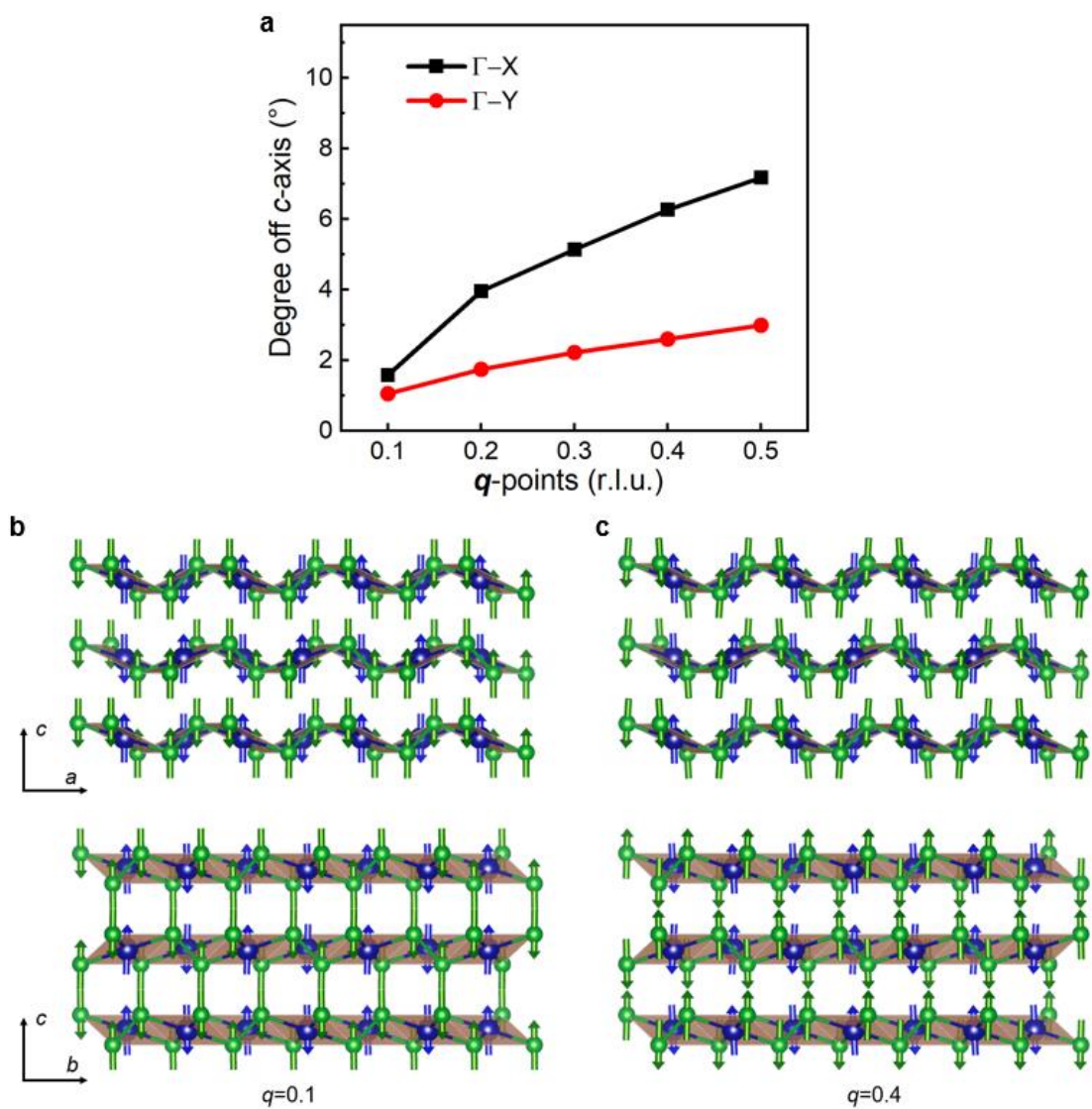

Figure S3. Deviation of the real part of the eigenvector from *c*-axis of the atoms in ZA modes. **a** Degree of the derivation of Pd atoms along Γ–X and Γ–Y directions. **b** and **c** are the vibration pattern of ZA mode along *a*- and *b*-axis at $q = 0.1$ and 0.4, respectively. Blue balls represent the Pd atoms, and green balls represent the Se atoms. Arrows are the displacement of each atom.

Figure S4a shows the percentage of each mode contributing to the lattice thermal conductivity along the individual directions. It is found that along *a*-axis, ZA, LA, TO$_{z1-z3}$, and ME contribute more to the thermal conductivity, while along *b*-axis, ZA, TA, TO$_{z1-z3}$, and LO contribute more to the thermal conductivity. The difference between each mode contribution to thermal conductivity along *a*-, *b*- and *c*-axis are shown in Figure S4b. It can be seen that 1$^{st}$–8$^{th}$ modes contribute more to the thermal conductivity. The mode contributions along *a*- and *b*-axis are higher than that along *c*-axis due to the weak interlayer van der Waals interactions. The difference between along *a*- and *b*-axis is mainly caused by 1st–8th modes discussed in detail in the main text.

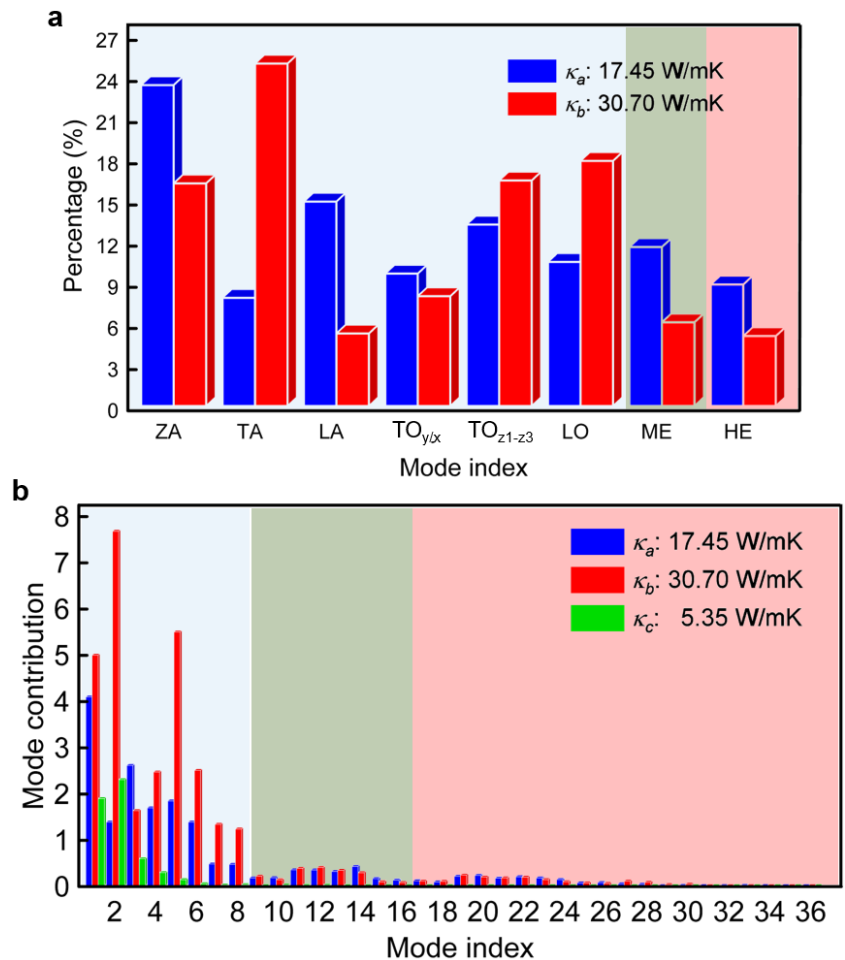

Figure S4. Mode contribution on the thermal conductivity of PdSe$_2$. **a** Percentage of the mode contribution of the LE phonons along a- and b-axis. **b** Mode contribution of each phonon branch on the thermal conductivity along *a*-, *b*-, and *c*-axis. Azure, dark-sea-green, and light-pink shadows represent the LE-, ME-, and HE-phonons, respectively.

Figure S5 shows the calculated energy-dependent phonon scattering rates of PdSe$_2$, where the HE-phonons possess high values. The four groups of scattering channels, including acoustic + acoustic → acoustic (*aaa*), acoustic + acoustic → optical (*aao*), acoustic + optical → optical (*aoo*), and optical + optical → optical (*ooo*), are separated to investigate the scattering phase space (inset). The scattering phase spaces of *aaa* and *aao* are much smaller than those of *aoo* and *ooo*. It suggests that the three-phonon scattering of LE phonons is limited to some extent.

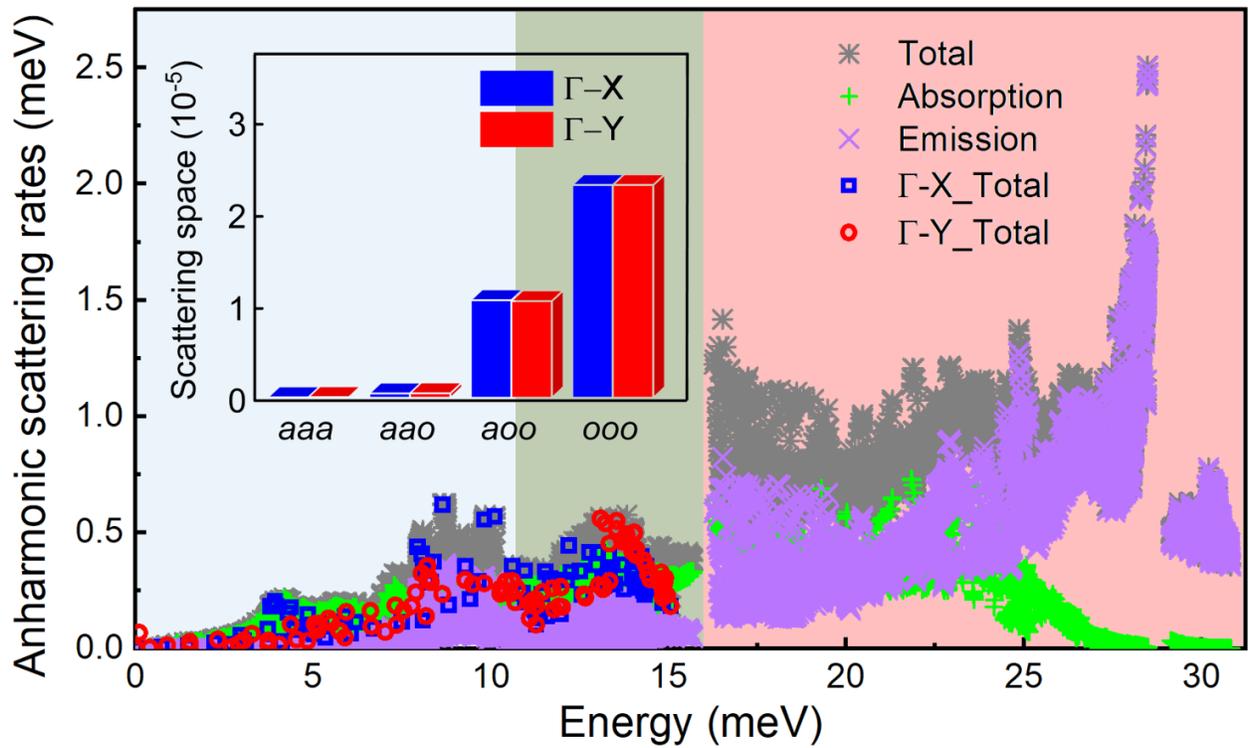

Figure S5. Calculated energy-dependent phonon scattering rates. Inset is the scattering phase space of *aaa*, *aao*, *aoo*, and *ooo* channels, indicating the weak scattering of the LE phonons. Azure, dark-sea-green, and light-pink shadows represent the LE-, ME-, and HE-phonons, respectively.

Figure S6 shows the frozen phonon potentials of the LE-optical modes along X–Γ–Y direction. It is found that all the LE-optical modes exhibit weak anharmonicity due to the near-quadratic profile.

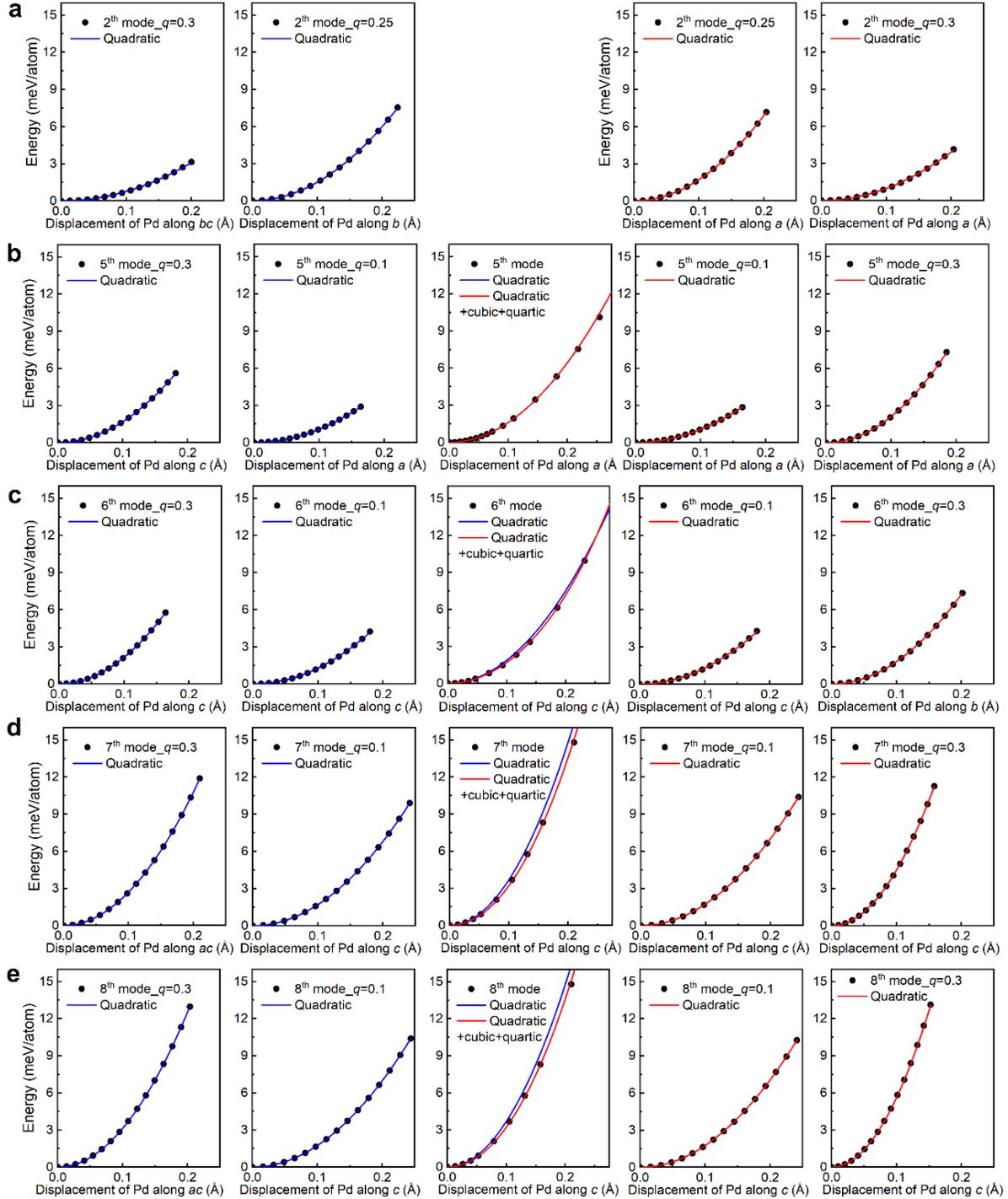

Figure S6. Frozen phonon potentials of the LE phonons along the *q*-path, indicating the harmonic behavior of these modes. **a**, **b**, **c**, **d**, and **e** are the potentials of the 2$^{nd}$ (TA), 5$^{th}$ (LO along Γ–X and TO$_x$ along Γ–Y) 6$^{th}$ (TO$_{Z1}$), 7$^{th}$ (TO$_{Z2}$), and 8$^{th}$ (TO$_{Z3}$) modes at *q*-points along X–Γ–Y direction from left to right, respectively.

Figures S7-S10 show the vibration patterns of TA, TO, LO, and $TO_{z1}$ modes, respectively. In TA, due to the avoided-crossing around $q = 0.3$ along Γ–X, there is a significant difference between the atomic vibrations in the zigzag and flat puckers. In TO, the atomic vibrations show typical features of the transverse optical mode. In the zigzag puckers, the atoms exhibit relatively large deviations from *b*-axis, while they exhibit slight deviations from *a*-axis in the flat puckers. In LO, it is found that the atoms vibrate nearly along the zigzag edge directions along a-axis, especially at high q-points, leading to a "bending-like" dispersive behavior along Γ–X. However, they vibrate nearly along *b*-axis, leading to a typical dispersive mode along Γ–Y, and thus the anisotropic phonon dispersion behavior. In $TO_{z1}$, the atoms show a typical feature of the out-of-plane vibration mode at $q = 0.1$. At $q = 0.3$, due to the avoided-crossing along both Γ–X and Γ–Y, the atoms show TA vibrations in the zigzag puckers while they show LA vibrations in the flat puckers. Therefore, the bucking structure induces significant differences in phonon dispersions between Γ–X and Γ–Y.

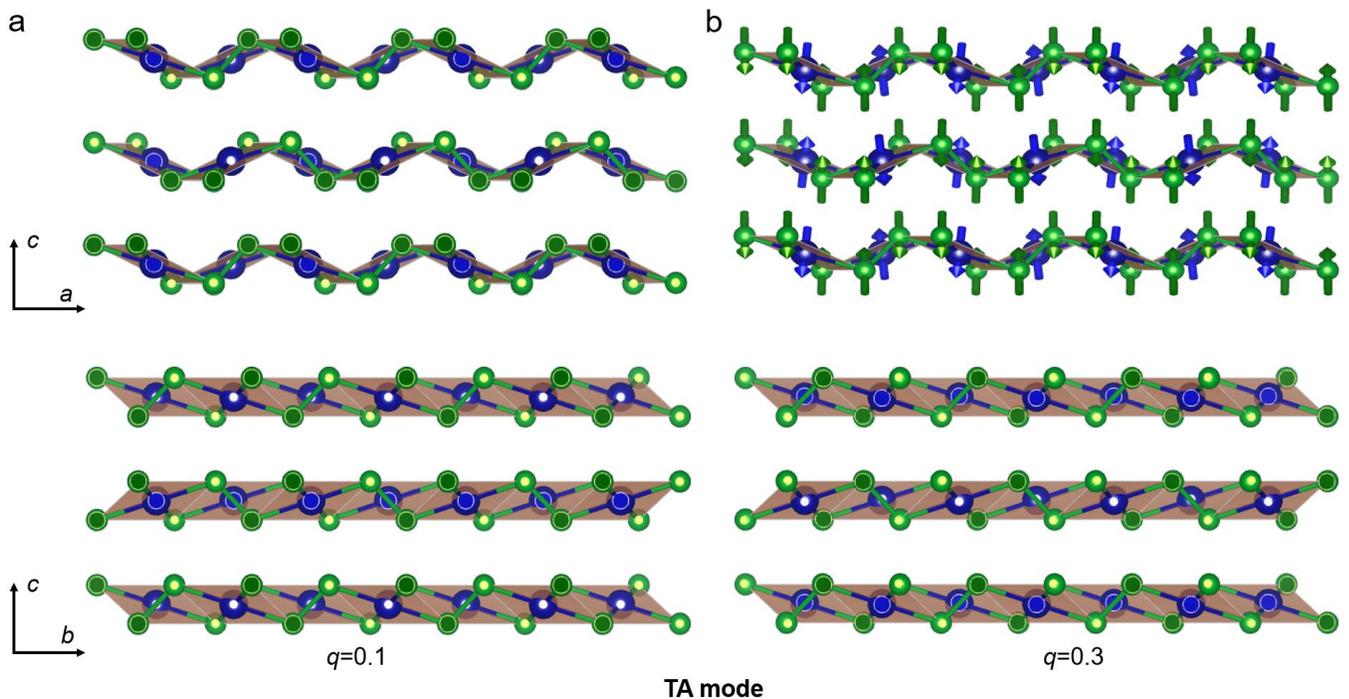

TA mode

Figure S7. Vibration patterns of TA mode along *a*- and *b*-axis at q = 0.1 (a) and 0.3 (b), respectively.

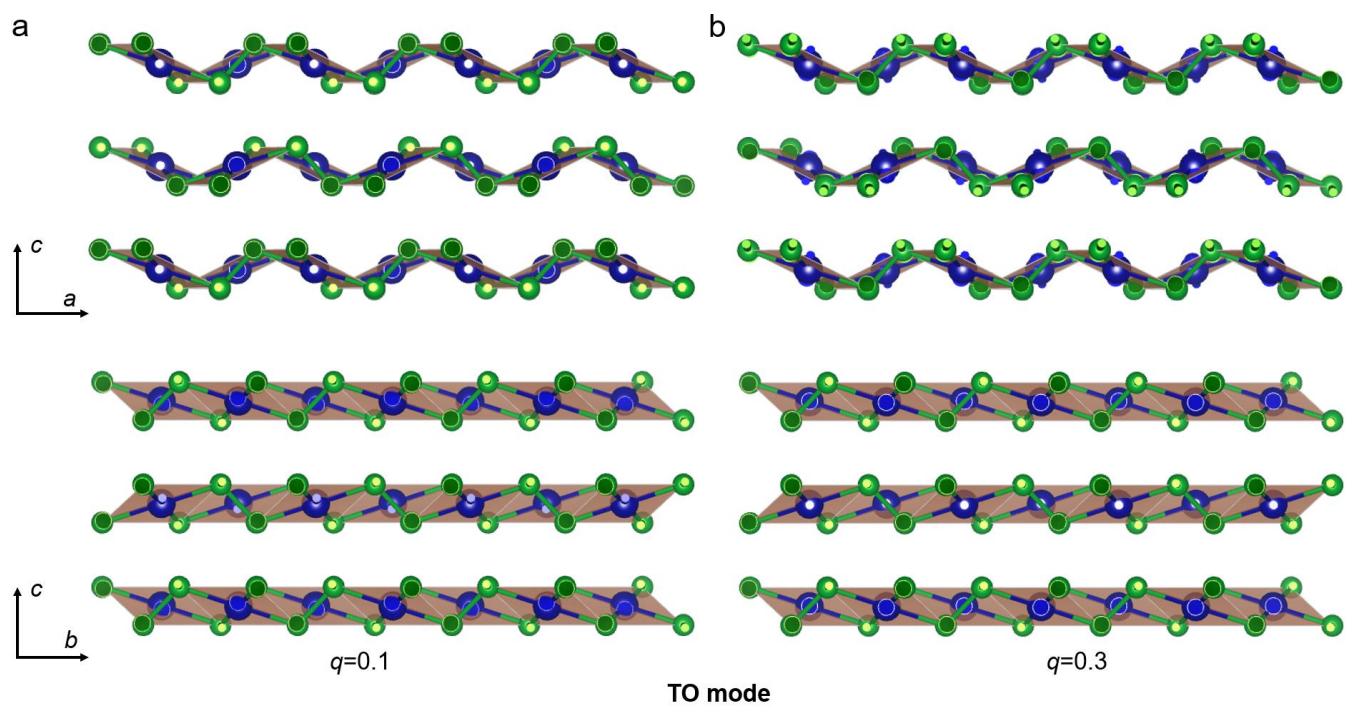

**TO mode**

Figure S8. Vibration patterns of TO mode along *a*- and *b*-axis at q = 0.1 (a) and 0.3 (b), respectively.

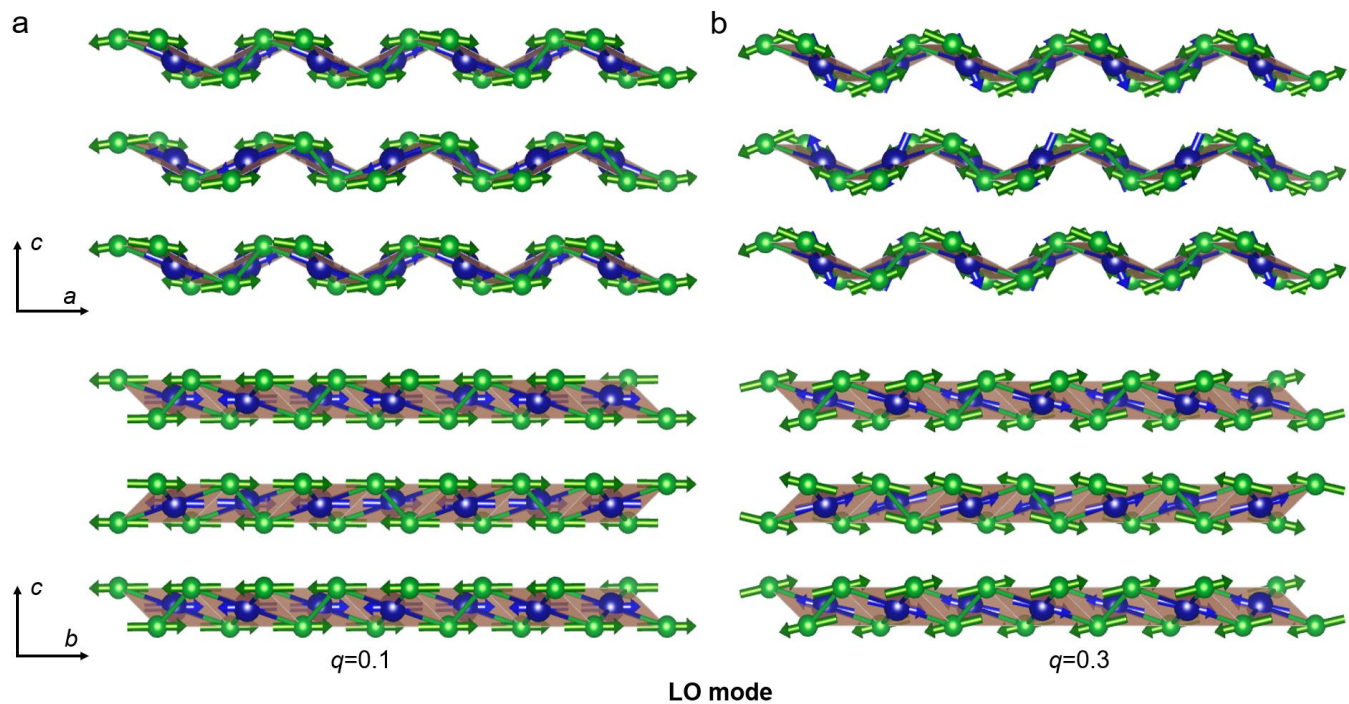

**LO mode**

Figure S9. Vibration patterns of LO mode along *a*- and *b*-axis at q = 0.1 (a) and 0.3 (b), respectively.

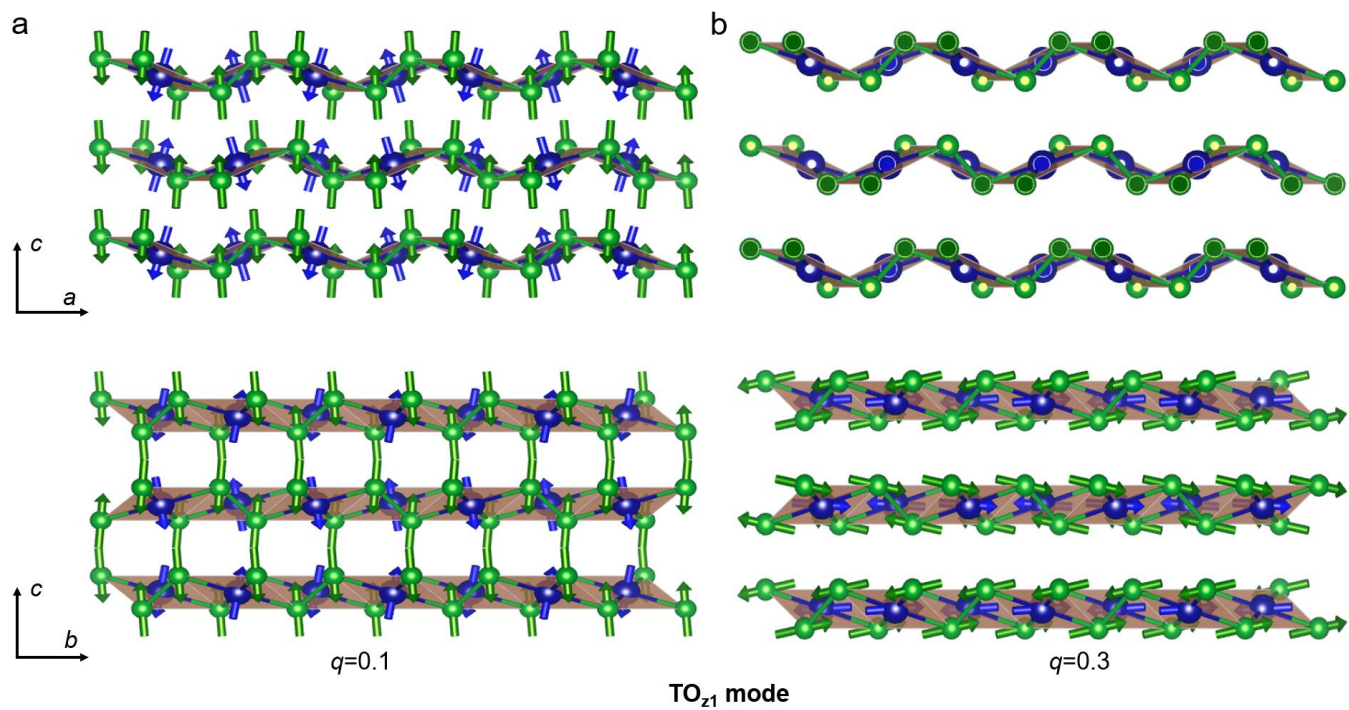

Figure S10. Vibration patterns of TO$_{z1}$ mode along *a*- and *b*-axis at q = 0.1 (a) and 0.3 (b), respectively.

Table SI lists the real parts of the eigenvectors of Pd atoms in ZA mode along *a*- (zigzag) and *b*-axis (planar) directions, based on which Figure S3 was obtained.

Table SI. Real parts of the eigenvectors of Pd atoms along zigzag and planar directions.

| *q*-points | Directions | Real part of the eigenvectors (Å) | | |
|---|---|---|---|---|
| | | X | Y | Z |
| 0.1 | Zigzag | -0.0013 | -0.0032 | 0.3185 |
| | Planar | -0.0003 | -0.0058 | 0.3168 |
| 0.2 | Zigzag | 0.0031 | 0.0053 | -0.3229 |
| | Planar | -0.0007 | 0.0096 | -0.3164 |
| 0.3 | Zigzag | 0.005 | 0.0059 | -0.334 |
| | Planar | 0.0002 | 0.0123 | -0.3175 |
| 0.4 | Zigzag | -0.0076 | -0.0053 | 0.3543 |
| | Planar | 0.0021 | -0.0145 | 0.3228 |
| 0.5 | Zigzag | -0.0371 | 0.0549 | 0.5261 |
| | Planar | -0.0114 | 0.022 | -0.4744 |